\documentclass[12pt]{article}
\usepackage{amsmath,amssymb}
\usepackage{graphicx,color}
\usepackage{cite}
\usepackage{bm}
\usepackage{dcolumn}

\def\be{\begin{equation}}
\def\ee{\end{equation}}
\def\bea{\begin{eqnarray}}
\def\eea{\end{eqnarray}}
\def\nn{\nonumber}

\newcommand{\Section}[1]{\section{#1}\setcounter{equation}{0}}

\begin{document}

\pagestyle{plain}

\def\e{{\rm e}}
\def\cs{\frac{1}{(2\pi\alpha')^2}}
\def\CV{{\cal{V}}}
\def\haf{{\frac{1}{2}}}
\def\tr{{\rm Tr}}
\def\"{\prime\prime}
\def\p{\partial}
\def\tphi{\tilde{\phi}}
\def\ttheta{\tilde{\theta}}
\def\a{\alpha}
\def\b{\beta}
\def\la{\lambda}
\def\barla{\bar{\lambda}}
\def\ep{\epsilon}
\def\hj{\hat j}
\def\nn{\nonumber}
\def\hn{\hat n}
\def\bz{\bar{z}}
\def\zk{{\bf{Z}}_k}
\def\h1{\hspace{1cm}}
\def\dd{\Delta_{[N+2k] \times [2k]}}
\def\ddbar{\bar{\Delta}_{[2k] \times [N+2k]}}
\def\u{U_{[N+2k] \times [N]}}
\def\ubar{\bar{U}_{[N] \times [N+2k]}}
\def\goes{\rightarrow}
\def\goal{\alpha'\rightarrow 0}
\def\ads2{AdS_2 \times S^2}
\def\ola{\overline {\lambda}}
\def\oep{\overline {\epsilon}}
\begin{titlepage}

\title{\bf 3-Form Flux Compactification of Salam-Sezgin Supergravity}
\author{Hamid Reza Afshar$^{1}$ \hspace{3mm} and \hspace{3mm} Shahrokh
Parvizi$^{1,2}$\\
{\small {\em 1. Department of Physics, Sharif University of Technology}} \\
{\small {\em P.O. Box 11155-9161, Tehran, IRAN}}\\
{\small {\em 2. Institute for Research in Fundamental Sciences (IPM),}} \\
{\small {\em P.O.Box 19395-5531, Tehran, Iran}} \\
{\small Emails:\hspace{2mm} h\underline{ }afshar@physics.sharif.ir,
\hspace{2mm} parvizi@theory.ipm.ac.ir}
 } \maketitle

\begin{abstract}
\noindent The compactification of 6 dimensional Salam-Sezgin model
in the presence of 3-form flux $H$ is investigated. We find a torus
topology for this compactification with two cusps which are the
places of branes, while at the limit of large size $L$ of the compact
direction we also obtain sphere topology.
This resembles the Randall-Sundrum I,II model. The
branes at one of the cusps can be chosen to be 3- and 4-branes which
fill our 4-dimensional space together with the fact that $H=0$ at this
position restores the Lorentz symmetry. This compactification also
provides an example for the so-called `time warp' solution, [0812.5107 [hep-th]].
According to a no-go theorem in $d\ne 6$, the time warp
compactification violates the null energy condition. While the
theorem is quiet for $d=6$, our model gives a time warp
compactification which satisfies the null energy condition. We also
derive the four dimensional effective Planck mass which is not
obvious due to the time warp nature of the solution.
\end{abstract}

\vspace{1cm} 
\vfill
\begin{flushbottom}
\vbox{
    \halign{#\hfil         \cr
        IPM/P-2009/021  \cr
           } 
      }   
\end{flushbottom}
\pagestyle{empty}
\end{titlepage}

\newpage
\Section{Introduction}

\vspace{4mm}

\noindent   In more than a decade, since the celebrated work of Randall-Sundrum \cite{Randall:1999ee,Randall:1999vf},
the warp compactification,  has been considered as a new approach to explain the hierarchy problem
in 4-dimensional space-time as a low energy limit of higher dimensional theories. This approach brought new
phenomenological results with more hopes
to find evidences for higher dimensional theories in a foreseeable future.

Long before the warp compactification idea, the six dimensional
gauged supergravity was studied by Salam-Sezgin in  \cite{Nishino:1984gk,Salam:1984cj,
RandjbarDaemi:1985wc,Nishino:1986dc}, as a simple model to obtain the supersymmetric vacua by
compactification to 4 dimensions. It has also interesting
applications in cosmological model building \cite{Maeda:1984gq,Maeda:1985es,Halliwell:1986bs,Papantonopoulos:2006uj}.
On the other hand, in another development \cite{Cvetic:2003xr},
it has been shown that this model can be derived from the string theory which strengthens its importance as a descendent
of a fundamental theory. In a modern view, the Salam-Sezgin supergravity is rich enough, while simple,
to provide the warp compactification including fluxes \cite{Carroll:2003db,Navarro:2003vw,Navarro:2003bf,
Aghababaie:2003wz,Aghababaie:2003ar,Gibbons:2003di,Burgess:2004dh}. In \cite{Gibbons:2003di}, it was found that
four dimensional Minkowski space solution is not only possible, but inevitable if one requires maximal symmetry in four
dimensions and compactness of internal space. Based on these features, it is worth to
work out its various warp compactifications.

The bosonic part of the model contains the metric, dilaton, a 2-form
$F_{(2)}$ and a 3-form $H_{(3)}$ as field strengths. In
\cite{Carroll:2003db,Navarro:2003vw,Navarro:2003bf,
Aghababaie:2003wz,Aghababaie:2003ar,Gibbons:2003di,Burgess:2004dh,Lee:2005az} a static warped solution has been found for
$H=0$ and $F\neq0$. A dynamical model was proposed in \cite{Copeland:2007ur}. For some recent developments see
\cite{Tolley:2006ht,Burgess:2006ds,Lee:2008pz}

In all of the case, so far $H$ has been set to zero. Beside technical reasons which make equations hard to solve when
$H$ is included, it is obvious that the presence of a 3-form in a 6-dimensional space can not support a symmetric
4-dimensional compactification. Nonetheless, we will see soon that the situation is not a disaster and one may find an
appropriate interpretation.

In this paper, we have considered a 4-dimensional compactification
with $H$ field which is extended along the 2-dimensional internal
space and the time direction. This kind of discrimination between
time and other non-compact spatial directions, may suggest its
application to cosmological models, however, here we restrict
ourselves to a static model and postpone the study of dynamical
solutions to future. Should we need a warp compactification, $H$
field configuration suggests the warp factors in time and spatial
directions should be different. This is what has been called `time
warp' recently in \cite{Gubser:2008gr}. The ratio of time and spatial warp
factor is the light speed which depends on the internal coordinate
by construction. There is a no-go argument in \cite{Gubser:2008gr}
according to which the internal space in time warped solutions can
not be compact, unless the null energy condition is violated.
Meanwhile the validity of this no-go theorem in  $d=6$ is under
query, and indeed our model provides a counterexample in which the
null energy condition can be satisfied even for the compact case.

We show that it is needed to solve equations in different patches and join them by Israel junction conditions
\cite{Israel:1966rt}. These conditions could be satisfied only when one introduces the branes at
joining positions \cite{Burgess:2008yx}. In this way we find out branes sitting at the middle and
two ends of the compact space. More explicitly, we consider a compact internal space with axial symmetry which
satisfies equations of motion in the interval $[0,L]$ for the radial coordinate, $z$, and then extended
to $[-L,0]$ interval with $L$ and $-L$ identified.
Thus we have a torus topology, with two cusps at 0 and $L$ which are the positions of branes. We consider minimal
number of branes and show that it is possible to introduce 3- and 4-branes filling our 4-dimensional
space where the 4-brane wrapped and 3-branes are smeared over the internal circle \cite{Burgess:2001bn}.
On the other side at $z=L$, in addition to 3- and 4-branes, we need 0-branes to satisfy the junction conditions
with time-space asymmetry. These 0-branes smeared over the world volume of the 4-brane.
This configuration makes it possible to have a 4-dimensional symmetric space at $z=0$. To ensure about this symmetry
we need to consider the behavior of $H$ field at $z=0$. Indeed $H$ is
discontinuous at this position, since branes act as a surface of polarized charges for the electrical $H$
field, so the $H$ field changes the sign while crossing the brane. The mean value of $H$ would be zero at $z=0$ which
together with the branes configuration restore the 4-dimensional lorentz symmetry at $z=0$.

At the first look, it may seem impossible to introduce an effective covariant 4-dimensional gravity, however, a fine
tuning of the parameters make it possible to obtain the effective Planck mass and 4-dimensional symmetry
in the linear approximation.

We organize the paper as follows. In the next section, equations of motion including the metric, dilaton and $H$ field
are solved. In section 3, we introduce the junction conditions and branes. These conditions also fix some of the
integration constants and we discuss the domain of independent parameters. The section 4 is devoted to discuss the
large $L$ limit where depending on the parameters, the internal azimuthal radius may diverge or shrink at large
$L$ to give new topologies. In section 5, we show the validity of the null energy condition. In section 6, the effective
four dimensional gravity is considered and the effective Planck mass is derived. We conclude in section 7.

\Section{Equations of motion and $H$-flux solution}

Let us start with the bosonic part of generalized Salam-Sezgin model with the following Lagrangian\cite{Nishino:1984gk,Salam:1984cj,
RandjbarDaemi:1985wc,Nishino:1986dc}:

\bea \label{lagrangian}
\frac{{\cal L}}{\sqrt{-g}}=\frac{1}{2\kappa^2}\left(-{\cal R}-\p_M\phi\p^M\phi \right)
- \frac{1}{4} \e^{-\phi} F_{MN}F^{MN}-\frac{1}{6}\e^{-2\phi} H_{MNP}H^{MNP}-
\frac{2g^2}{\kappa^4} \e^{\phi}
\eea

where capital latin indices are six dimensional indices, $\phi$ is the dilaton, $F$ and $H$ are 2 and 3-form fields.
Equations of motion follows as,

\bea
-{\cal R}_{MN}&=&\p_M\phi\p_N\phi+\kappa^2 \e^{-\phi} \left(F_{MN}^2- \frac{1}{8}F^2
G_{MN} \right) \nn\\
&&+ \haf\kappa^2 \e^{-2\phi} \left(H_{MN}^2- \frac{1}{6}H^2 G_{MN} \right)
+\frac{g^2}{\kappa^2} \e^{\phi} G_{MN}\nn
\eea

\bea
&&  \Box \phi + \frac{\kappa^2}{6}\e^{-2\phi} H_{MNP}H^{MNP}
+\frac{\kappa^2}{4} \e^{-\phi} F_{MN}F^{MN}-\frac{2g^2}{\kappa^2} \e^{\phi} =0 \nn\\
&&  D_M\left(\e^{-2\phi}H^{MNP}\right)=0  \nn\\
&&  D_M\left(\e^{-\phi} F^{MN}\right) + \e^{-2\phi}H^{MNP}F_{MP} =0
\eea

To solve the above equations, we consider compactification to 4-dimension with axial symmetry in the
internal space. Since we are looking for static solutions, we take all fields to be dependent on the
internal radial coordinate $\eta$ as in the following ansatze,

\bea \label{metric}
ds^2= -\e^{2w(\eta)} dt^2 +\e^{2a(\eta)} \delta_{ij} dx^i dx^j +\e^{2v(\eta)}d\eta^2 +\e^{2b(\eta)} d\theta^2  \nn\\
F=0\;, \;\;\;\;\;\; \e^{\phi}=\e^{\phi(\eta)}, \;\;\;\;\;\; H= h'(\eta) dt \wedge d\theta \wedge d\eta \;. \;\;\;\;
\eea

For dimensional convenience we assume $\theta$ has length of dimension with  $0 \le \theta \le L_\theta$.
since $H$ extensions distinguish the time from other spatial noncompact coordinates, we have included two different
warp factors $\e^{2w}$ and $\e^{2a}$ in the metric. Now the equations read as,

\bea \label{Maxwell}
({\rm Maxwell})&&  h^{\"}+(3a'-w'-v'-b'-2\phi')h'=0   \\
\label{dilaton}
({\rm Dilaton})&&  \phi^{\"}+(3a'+w'-v'+b')\phi'-\kappa^2 h'^2 \e^{-2(w+b+\phi)}-
\frac{2 g^2}{\kappa^2} \e^{2v+b}=0
\eea
\bea
\label{teinstein}
(tt\;{\rm Einstein})&&   w^{\"}+(w' +3a'-v'+b')w'- \frac{\kappa^2h'^2}{2} \e^{-2(w+b+\phi)} +
\frac{g^2}{\kappa^2}\e^{2v+\phi}=0  \nn\\
\label{ieinstein}
(ii\;{\rm Einstein})&&   a^{\"}+(w' +3a'-v'+b')a'+\frac{\kappa^2h'^2}{2} \e^{-2(w+b+\phi)}
+\frac{g^2}{\kappa^2}\e^{2v+\phi}=0   \nn\\
\label{thetaeinstein}
(\theta\theta\;{\rm Einstein})&&   b^{\"}+(w' +3a'-v'+b')b'-\frac{\kappa^2h'^2}{2} \e^{-2(w+b+\phi)}
+\frac{g^2}{\kappa^2}\e^{2v+\phi}=0   \nn\\
\label{etaeinstein}
(\eta\eta\;{\rm Einstein})&&   w^{\"}+3a^{\"}+b^{\"}+w'^2+3a'^2+b'^2+\phi'^2-(w'+3a'+b')v'  \nn\\
&& -\frac{\kappa^2h'^2}{2} \e^{-2(w+b+\phi)}+\frac{g^2}{\kappa^2}\e^{2v+\phi}=0  \nn\\
\eea

To solve these equations we can use the gauge freedom in choosing coordinate $\eta$  such that,

\bea\label{gauge}
(w' +3a'-v'+b')=0
\eea
Then suitable combinations of (\ref{Maxwell})-(\ref{etaeinstein}) give,
\bea
\label{hprime}
h'(\eta)&=&\pm q \e^{2x} \nn\\
\label{w1}
w(\eta) &=& \frac{y+x}{4}+(2\la_3+\la_4)\eta  \nn\\
\label{a1}
a(\eta) &=& \frac{y-x}{4}+\left(\frac{-\la_3}{3}\right)\eta  \nn\\
\label{v1}
v(\eta) &=& \frac{5y-x}{4}+\la_3\eta  \nn\\
\label{b1}
b(\eta) &=& \frac{y+x}{4}-\la_4\eta  \nn\\
\phi(\eta) &=& \frac{x-y}{2}-2\la_3\eta \nn\\
\eea
with $q$ a real
positive number and $x(\eta)$ and $y(\eta)$ can be found from, \bea
x'^2-2\kappa^2q^2\e^{2x} &=& \la_1^2   \nn\\
y'^2+\frac{4g^2}{\kappa^2}\e^{2y}&=& \la_2^2 \eea and $\la_i$'s are
integration constants which are not independent and satisfy,
\bea \label{constraint}
\la_2^2=\la_1^2+2\left(\la_3+\la_4 \right)^2+\frac{16}{3} \la_3^2
\eea
The general solutions of these equations are:

\bea\label{x}
\e^{-x}&=&\frac{\sqrt{2}\kappa q}{\la_1}  f(\la_1(\eta-\eta_1))  \nn\\
\e^{-y}&=&\frac{2g}{\kappa \la_2} \cosh (\la_2(\eta-\eta_2))  \nn\\
f(\eta)&=& \left\{
  \begin{array}{lc}
    \pm \sinh (\eta)  & \la_1^2 > 0  \\
    \pm \eta  & \la_1^2 = 0 \\
    \pm \sin (\eta) & \la_1^2 < 0 \\
  \end{array}
\right.
\eea
$\lambda_2$ is positive, since $g$, $\kappa$, $\e^{-y}$ are
non-negative values. To ensure that $\e^{-x}$ for all $\eta$ is a
non-zero positive real number we can construct its solution as:
\bea \e^{-x}&=& \left\{
  \begin{array}{cc}
     \frac{\sqrt{2}\kappa q}{\la_1} \;f(\la_1(\eta-(\eta_1-\varepsilon)) & \eta>\eta_1  \nn\\
    -\frac{\sqrt{2}\kappa q}{\la_1} \;f(\la_1(\eta-(\eta_1+\varepsilon)) & \eta<\eta_1  \nn\\
  \end{array}
\right.
\eea
where $\varepsilon>0$. If we change the coordinate as follows,
\bea
   z&=&\la_2(\eta-\eta_1)\nn\\
 z_1&=&\la_2\varepsilon\nn\\
 z_2&=&\la_2(\eta_2-\eta_1)\nn\\
 \la&=&\frac{\la_1}{\la_2}\nn
\eea
and use the absolute value, we can construct an even solution with respect to $z=0$. So we find
\bea \e^{-x}&=&\frac{\kappa \widetilde{q}}{\la}  f(\la(|z|+z_1))  \nn\\
\e^{-y}&=&\frac{\widetilde{g}}{\kappa } \cosh (|z|-z_2)
\eea
where
\bea \widetilde{q}=\frac{\sqrt{2}q}{\la_2}
\;\;\;\;,\;\;\;\; \widetilde{g}=\frac{2g}{\la_2} .
\eea

The constraint (\ref{constraint}) can be written as:
\bea \label{constraint1} 1 = \la^2 + 2(
\widetilde{\la}_3 + \widetilde{\la}_4 )^2 + \frac{16}{3}
\widetilde{\la}_3^2 \eea where \bea
\widetilde{\la}_3=\frac{\la_3}{\la_2} \;\;\;\;,\;\;\;\;
\widetilde{\la}_4=\frac{\la_4}{\la_2} \;\;\;\;,\;\;\;\;
\la=\frac{\la_1}{\la_2}\nn\;.
\eea

So far we have derived general solutions to the equations of motion including integration constants.
To fix these constants, we need appropriate boundary conditions or physically interesting special cases.
We deal with these conditions in the following sections.

\begin{figure}
 \begin{center}
     \begin{tabular}{c c}
       Finite $L$ & Large $L$  \\
       \includegraphics[scale=.4]{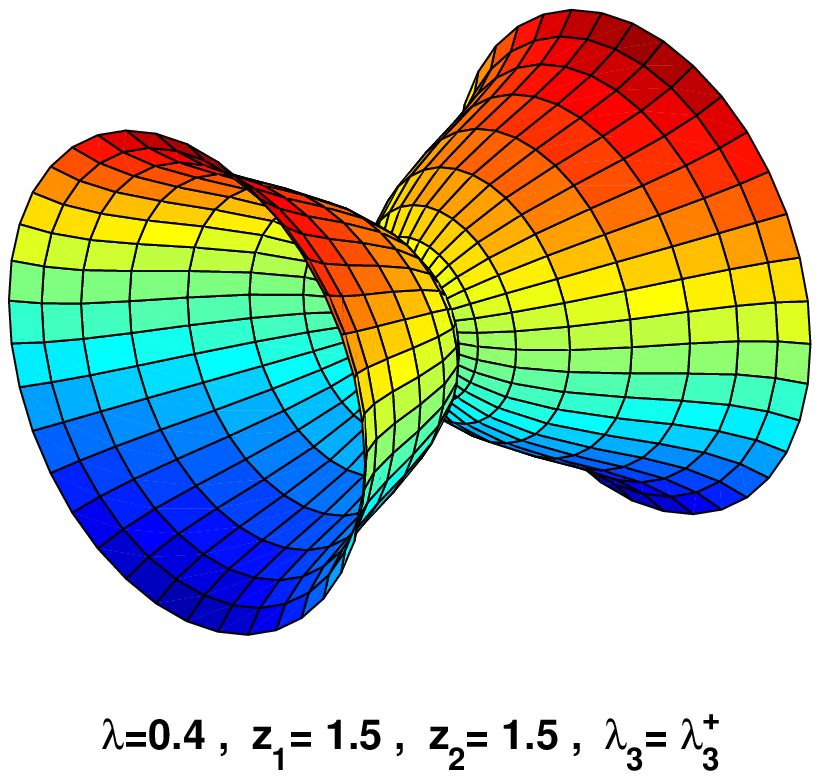} & \includegraphics[scale=.4]{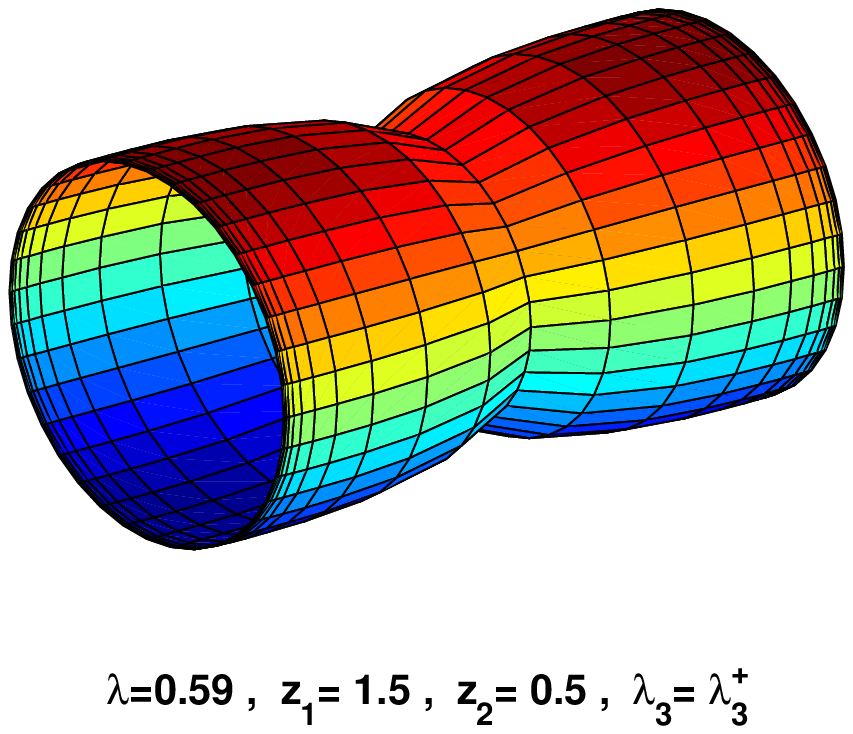} \\
       (\textit{a}) & (\textit{b}) \\
       \includegraphics[scale=.4]{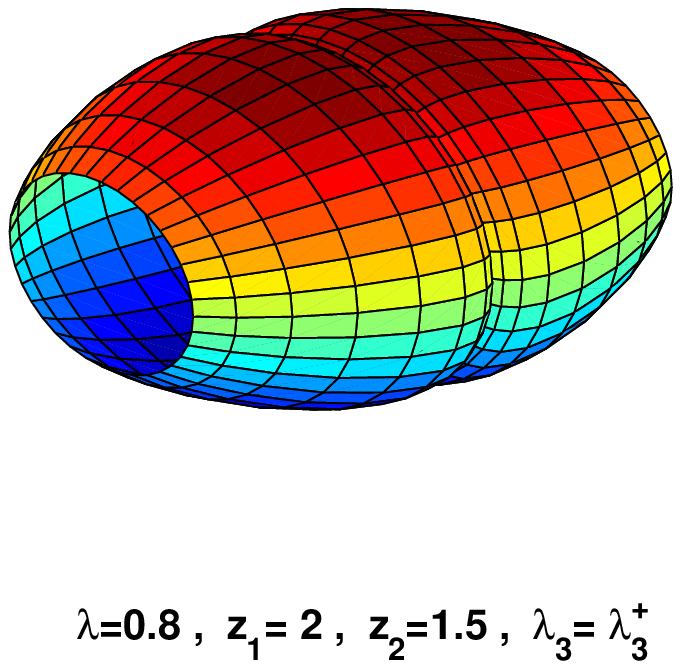} & \includegraphics[scale=.4]{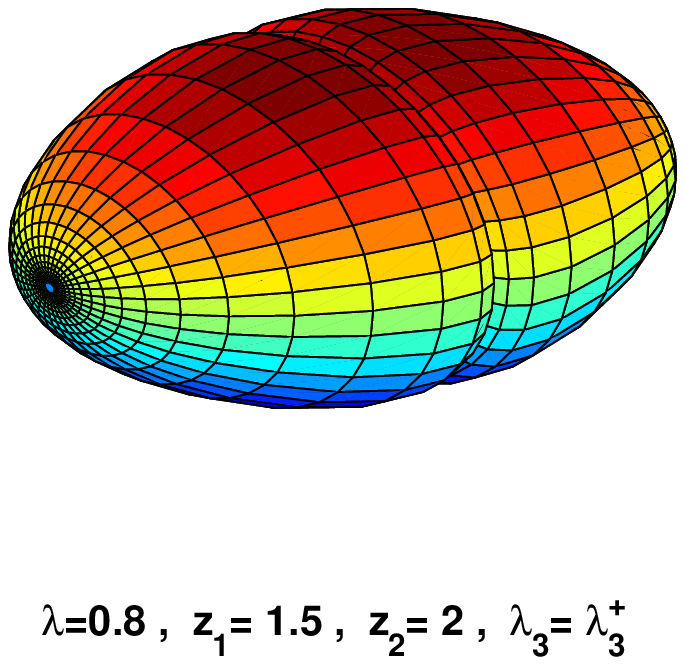} \\
       (\textit{c}) & (\textit{d}) \\
       \includegraphics[scale=.4]{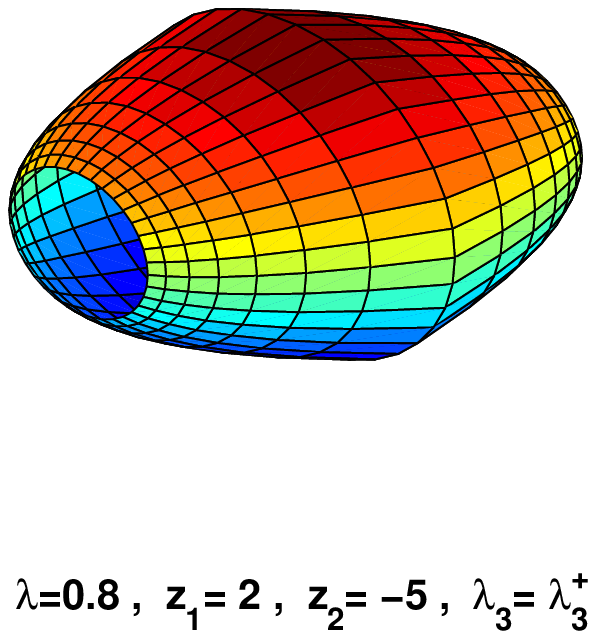} & \includegraphics[scale=.4]{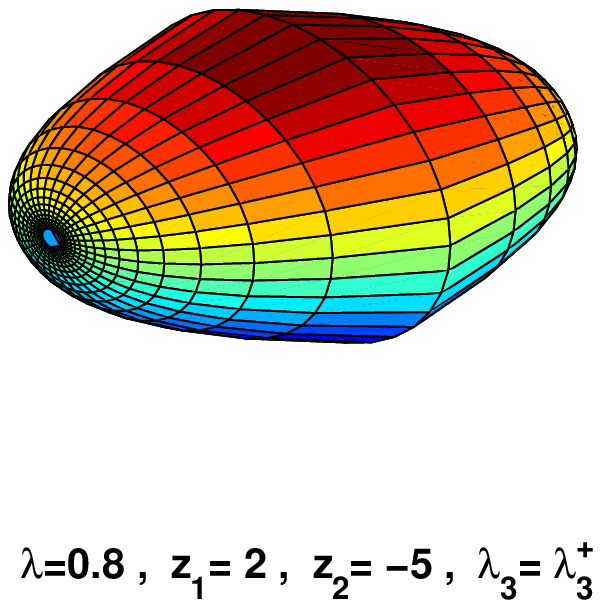} \\
       (\textit{e}) & (\textit{f}) \\
       \includegraphics[scale=.4]{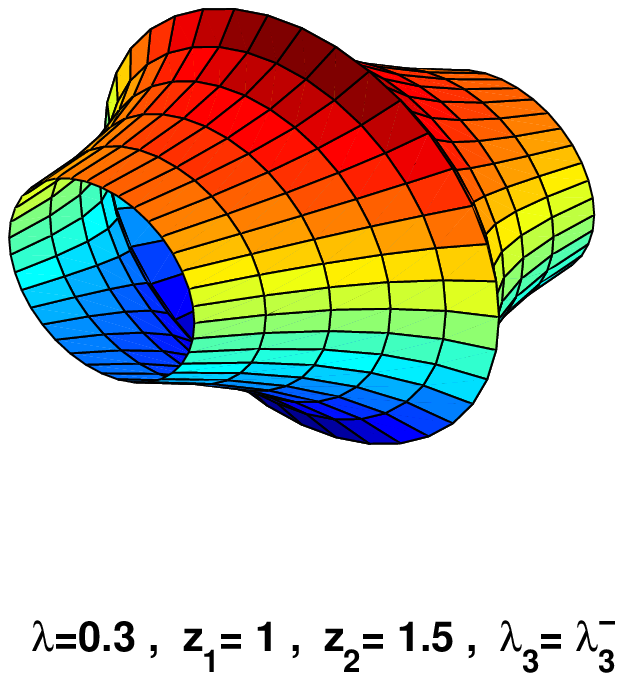} & \includegraphics[scale=.4]{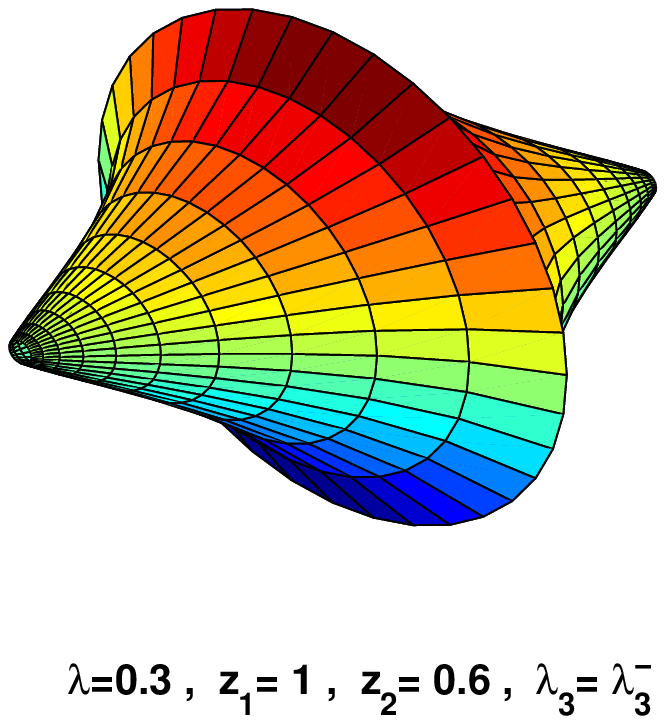} \\
       (\textit{g}) & (\textit{h}) \\
     \end{tabular}
 \end{center}
\caption{The shapes of internal space for various parameters
. The axial direction is the
$\rho$-axis. $\beta < 0$ for (\textit{a}), $\beta=0$ for (\textit{b}) and $\beta>0$ for others.}\label{FIG}
\end{figure}

\Section{Branes and Israel junction conditions}
In this section we study the global
aspects of the above solution. Firstly as stated below (\ref{x}), we
should keep the exponential functions in the metric to be positive everywhere
and this indicates that the above solutions can not be valid globally, we need to
cut and join them in different patches appropriately. This has
already been done at $z=0$. Also trying to find  a compact internal space,
we take the $z$ direction to be compact in some
interval $[-L,L]$ with periodic boundary conditions\footnote{The Euler character can be calculated,
\bea
\chi = \frac{1}{4 \pi} \int_Y \sqrt{g} R^{(2)} d^2y + \frac{1}{2 \pi}\int_{\p Y} K ds  \nn
\eea
where $K=g^{\theta\theta}K_{\theta\theta}$ is the geodesic curvature on the boundaries.
Then,
\bea
\chi&=&2\frac{L_\theta}{2 \pi} \left[\int_0^L \e^{b-v} \left(b''+b'^2-b'v'\right)dz
- b'\e^{b-v}|_{0} + b'\e^{b-v}|_{L}\right] = 0   \nn
\eea
This shows that the internal space is generically a torus. The Large $L$ limit may cause a cycle shrinks as
can be seen in cases \textit{d}, \textit{f} and \textit{h} of figure~\ref{FIG} .}.
We will study the noncompact limit ($L \goes \infty$) later. Indeed the solution
set in the previous section is valid for each segment of $(-L,0)$
and $(0,L)$. Thus we only need to match different patches by Israel
junction conditions. We know that these conditions ensure the
continuity of the solutions and relate the derivative discontinuities to
possible  brane tensions. So we expect there might be some branes
sitting at $z=0$ and/or $z=L$.

The Israel junction conditions relate the jump in the derivatives of
the metric to the branes tension sitting at $z=z_0$ as follows,
\bea\label{jun1}
[K_{mn}-K\hat{g}_{mn}]_{z_0}+\kappa^2t_{mn}&=&0
\eea
where [$f(z)]_{z_0}$ means
\bea
[f(z)]_{z_0}:=\lim_{\epsilon \goes 0^+} \left(f(z_0+\epsilon)-f(z_0-\epsilon) \right)\nn
\eea
and $K_{mn}$ is the extrinsic curvature of constant proper radius $\rho$ which is introduced
in the following form of the metric:
\bea
ds^2=d\rho^2+\hat{g}_{mn}dx^mdx^n \;.
\eea
Then the extrinsic curvature is $K_{mn}=\frac{1}{2}\partial_{\rho}\hat{g}_{mn}$.
The brane stress energy $t^{mn}$ is given by
\bea
t^{mn}\equiv\frac{2}{\sqrt{-\hat{g}}}\frac{\delta S_{brane}}{\delta
\hat{g}_{mn}}
\eea
In our case because the 4D maximal symmetry has
been broken out, it is impossible to interpret the 4-brane stress
tensor as being due to a pure tension. But we can be hopeful to find
it at least along one of the branes at e.g. $z=0$:
\begin{eqnarray}\label{jun2}
  t_{\mu\nu} &=& \la_2 T \;\hat{g}_{\mu\nu} \nn\\
  t_{\theta\theta} &=& \la_2 T_4 \;\hat{g}_{\theta\theta}
\end{eqnarray}
where  $T=T_4+\widetilde{T}_{3}$  with
$\widetilde{T}_{3}=\frac{T_3}{L_\theta}$.  $\la_2$ is inserted  for
later convenience.  These are the configuration of the stress energy
tensors of a four-brane wrapping the internal circle and a
three-brane which is smeared over the internal circle.  This
situation can't be satisfied for the other side at $z=L$
simultaneously, so in the most general form, the stress energy
tensors at $z=L$ is taken to be:
\begin{eqnarray}
  t_{00} &=& \la_2(\widetilde{T}_{L0}+T_{L4}+\widetilde{T}_{L3})\hat{g}_{00}\nn\\
  t_{ij} &=& \la_2(T_{L4}+\widetilde{T}_{L3}) \hat{g}_{ij}\nn\\
\label {jun5}  t_{\theta\theta} &=& \la_2 T_{L4} \hat{g}_{\theta\theta}
\end{eqnarray}
where in addition to 3 and 4-branes, we have considered 0-branes
at $L$ smeared over all spatial direction except for $z$ direction. The
{\it tilde} over the tensions shows they are the density of smeared
tensions, i.e., $\widetilde{T}_{L0}=T_{L0}/Vol_4$ and
$\widetilde{T}_{L3}=T_{L3}/L_\theta$.

Plugging our solution to the junction conditions (\ref{jun1}), and after appropriate combinations,
we obtain the following conditions at $z=0,L$:
\bea\label{w'minusa'}
{[}a'(z)-w'(z){]}_{z=0} &=& 0  \nn\\
{[}b'(z)-a'(z){]}_{z=0} &=& \kappa^2 \e^{v(0)} \widetilde{T}_{3}   \nn\\
{[}3a'(z)+w'(z){]}_{z=0} &=& \kappa^2\e^{v(0)} T_4  \nn\\
{[}a'(z)-w'(z){]}_{z=L} &=&  \kappa^2 \e^{v(L)}  \widetilde{T}_{L0}  \nn\\
{[}b'(z)-a'(z){]}_{z=L} &=& \kappa^2 \e^{v(L)}  \widetilde{T}_{L3}   \nn\\
{[}3a'(z)+w'(z){]}_{z=L} &=& \kappa^2 \e^{v(L)} T_{L4} \;.\nn\\
\eea

Let us consider the above conditions on the \textit{sinh} solution. The
\textit{sine} and \textit{linear} solutions can be derived by taking $\la\goes
i\la$ and $\la \goes 0$, respectively. Firstly, write the solution as,

\bea
\e^{-x(z)}&=&\frac{\kappa \widetilde{q}}{\la}\sinh\left( \lambda\, \left( |z|+z1 \right)  \right)\theta \left( L-|z|\right) \nn \\
\e^{-y(z)}&=&\frac{\widetilde{g}}{\kappa }  \cosh \left(  |z|-z2   \right)    \theta \left( L-|z| \right) \nn\\
w(z) &=& \frac{y+x}{4}  + (2\widetilde{\la}_3+\widetilde{\la}_4)(|z|+z_3)  \theta \left( L-|z| \right) \nn\\
a(z) &=& \frac{y-x}{4}  -\frac{\widetilde{\la}_3}{3} (|z|+z_3)  \theta\left( L-|z| \right) \nn\\
v(z) &=& \frac{5y-x}{4} +\widetilde{\la}_3(|z|+z_3)  \theta \left( L-|z| \right) \nn\\
b(z) &=& \frac{y+x}{4}  -\widetilde{\la}_4(|z|+z_3)  \theta \left( L-|z| \right) \nn\\
\phi(z) &=& \frac{x-y}{4} -2\widetilde{\la}_3(|z|+z_3)  \theta\left( L-|z| \right)\nn\\
\eea
where $\theta(z)$ is the Heaviside step function:

\bea
\theta(z)= \left\{  \begin{array}{cc}    1  & z > 0  \\    0  & z < 0 \\  \end{array}\right.\nn
\eea

The solutions are continuous at $z=0,\pm L$ and we demand them to be periodic with respect to $2L$ shift.

The first condition of (\ref{w'minusa'}) gives the following constraint,

\bea
3 \lambda \coth(\lambda z_1) = 14 \widetilde{\lambda}_3 + 6 \widetilde{\lambda}_4
\eea
from which together with (\ref{constraint1}) we obtain two constants
$\widetilde{\la}_3$ and $\widetilde{\la}_4$,

\bea \label{lambda34}
\widetilde{\la}_3^{\pm}&=& \frac{3}{20}c\lambda \pm \frac{3}{40}\sqrt{-6c^2\lambda^2-20\lambda^2+20}   \\
\widetilde{\la}_4^{\pm} &=& \frac{3}{20}c\lambda \mp \frac{7}{40}\sqrt{-6c^2\lambda^2-20\lambda^2+20}
\eea
where $c= \coth{\la z_1}$ and the reality condition for $\widetilde{\la}_3$ and
$\widetilde{\la}_4$ imposes the following inequality,

\bea
 z_1 \ge \frac{1}{\la} \log{\sqrt{k+\sqrt{k^2-1}}}
\eea
where $k=(10-7\la^2)/(10-13\la^2)$.

\begin{figure}
 \begin{center}\begin{tabular}{c c}
                 \includegraphics[scale=.4]{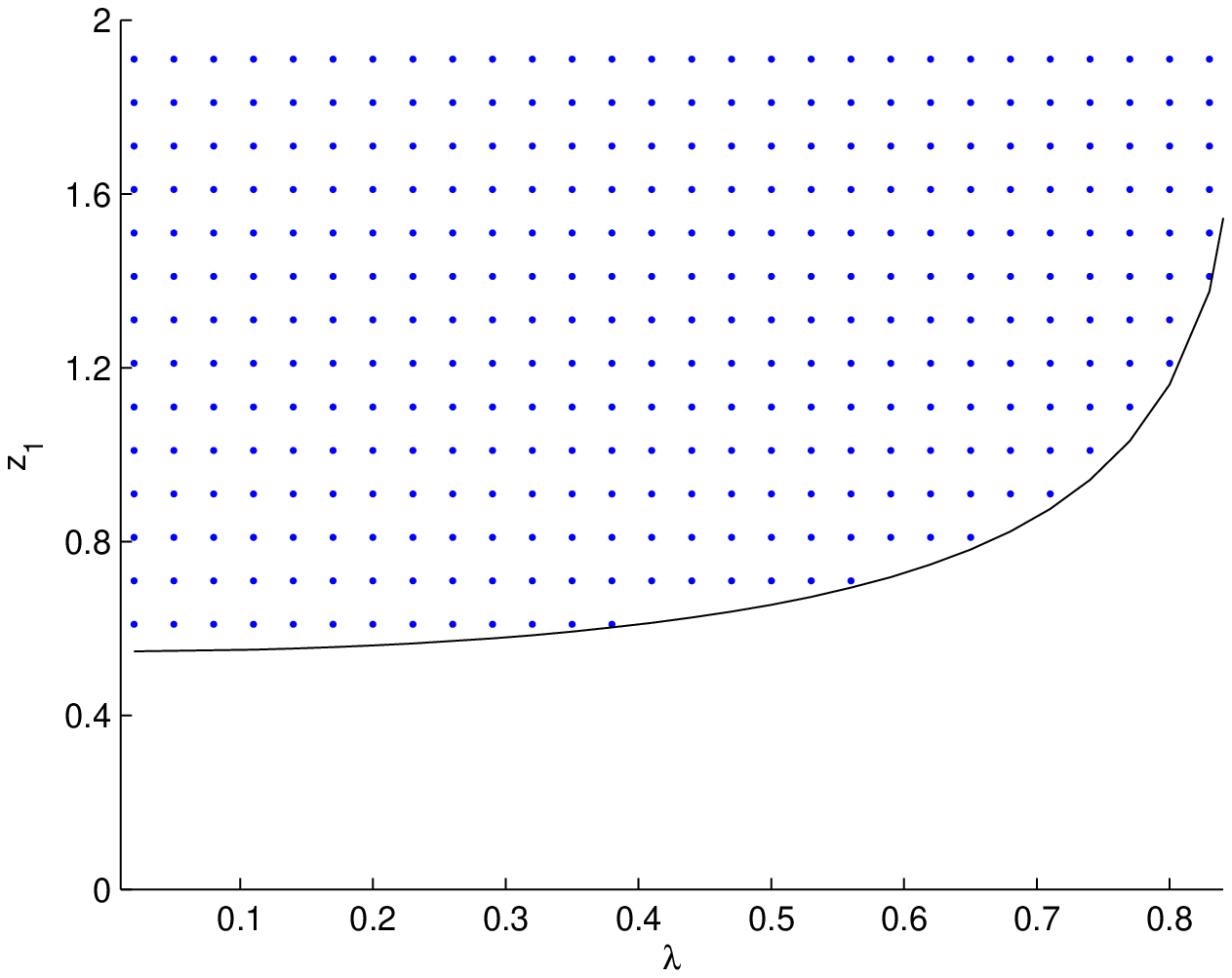} & \includegraphics[scale=.4]{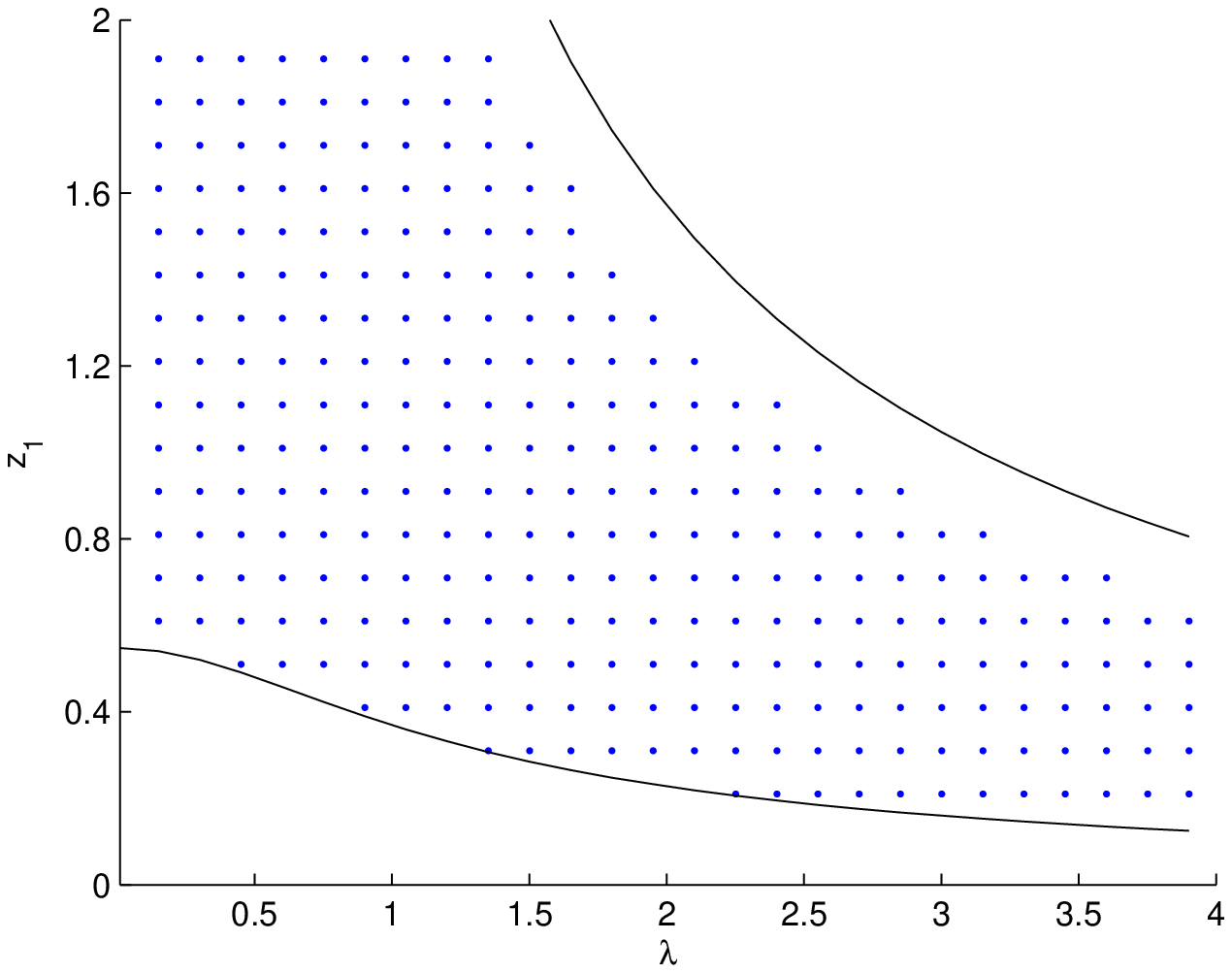} \\
                 \textit{a}) $\textit{sinh}$ solution &  \textit{b}) $\textit{sine}$ solution \\
               \end{tabular}
 \end{center}
\caption{The dotted regions are permitted values of $\la$ and $z_1$ for which we have real
parameters $\widetilde{\la_3}$ and $\widetilde{\la_4}$.
In \textit{a} the region is asymptote to maximum $\la$ at $\sqrt{10/13}$. In \textit{b} the upper curve shows an upper bound as
$z_1 < \pi/\la$. Considering finite $L$ the region $z_1 < \pi/\la-L$ gets smaller.}\label{reallambda}
\end{figure}

Similarly for $sine$ and $linear$ solutions where $\la\goes i\la$
and $\la \goes 0$, respectively, we find the following regions in $\la-z_1$ plane:

\begin{center}
\begin{tabular}{l r}
  $z_1 \ge \frac{1}{\la} \sin^{-1}\left( \sqrt{\frac{3\la^2}{13\la^2+10}} \right)$ & {\rm sine solution} \\
  $z_1 \ge \sqrt{\frac{3}{10}}$  & {\rm linear solution} \\
\end{tabular}
\end{center}
For the \textit{sine} case we require that \textit{sine} to be positive which gives $0< \la(z+z_1) <\pi$,
thus  $(L+z_1)< \pi/\la$.
The permitted regions in $\la-z_1$ plane are drown in figure~\ref{reallambda}.

From the other five junction conditions in (\ref{w'minusa'}) we derive the brane tensions,

\bea
\kappa^2 T &=& \left(\frac{8}{3}\widetilde{\la}_3+2 \tanh(z_2)\right)\e^{-v(0)}  \nn\\
\kappa^2 T_4 &=& \left(\frac{20}{3}\widetilde{\la}_3 +4\widetilde{\la}_4 +2 \tanh(z_2)\right)\e^{-v(0)}  \nn\\
\kappa^2 T_{L0} &=& \left(\frac{14}{3}\widetilde{\la}_3+2\widetilde{\la}_4-\la\coth(\la(L+z_1))\right)\e^{-v(L)}  \nn\\
\kappa^2 T_{L4} &=& \left(-2\widetilde{\la}_3 -2\widetilde{\la}_4-\la\coth(\la(L+z_1))+2 \tanh(L-z_2)\right)\e^{-v(L)}  \nn\\
\kappa^2 \widetilde{T}_{L3} &=& \left(-\frac{2}{3}\widetilde{\la}_3+2\widetilde{\la}_4+\la\coth(\la(L+z_1)) \right)\e^{-v(L)} \nn\\\eea
where
\bea
\e^{-4v(0)} &=& \frac{\widetilde{g}^5\la}{\kappa^6\widetilde{q}}\;\frac{\cosh^5(z_2)}{\sinh(\la z_1)} \e^{-4\widetilde{\la}_3 z_3}  \nn \\
\e^{-4v(L)} &=& \frac{\widetilde{g}^5\la}{\kappa^6 \widetilde{q}}\;\frac{\cosh^5(L-z_2)}{\sinh(\la (L+z_1))} \e^{-4\widetilde{\la}_3(L+z_3)}
\eea

\begin{figure}
 \begin{center}
                 \begin{tabular}{c c}
                   \includegraphics[scale=.4]{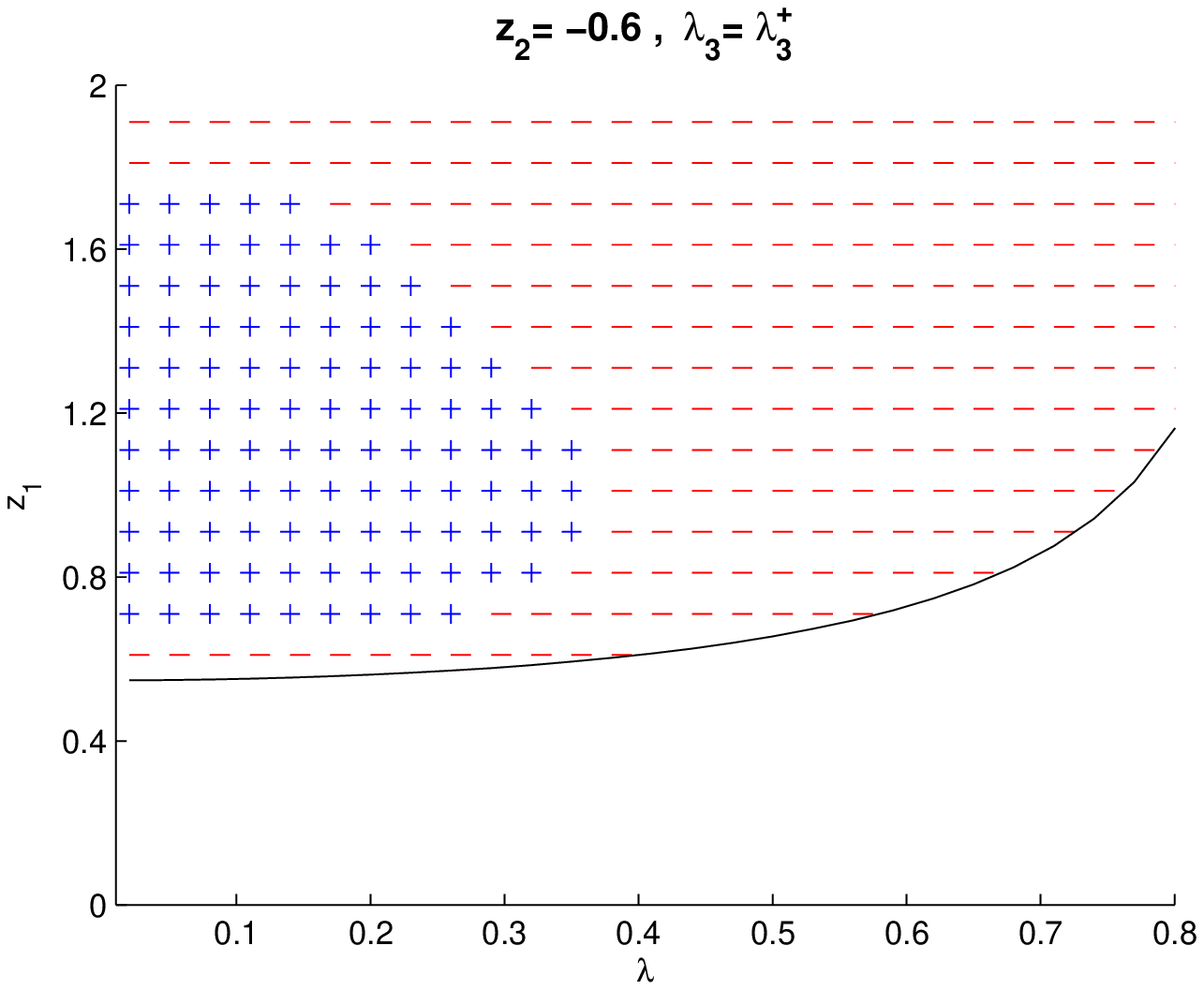} & \includegraphics[scale=.4]{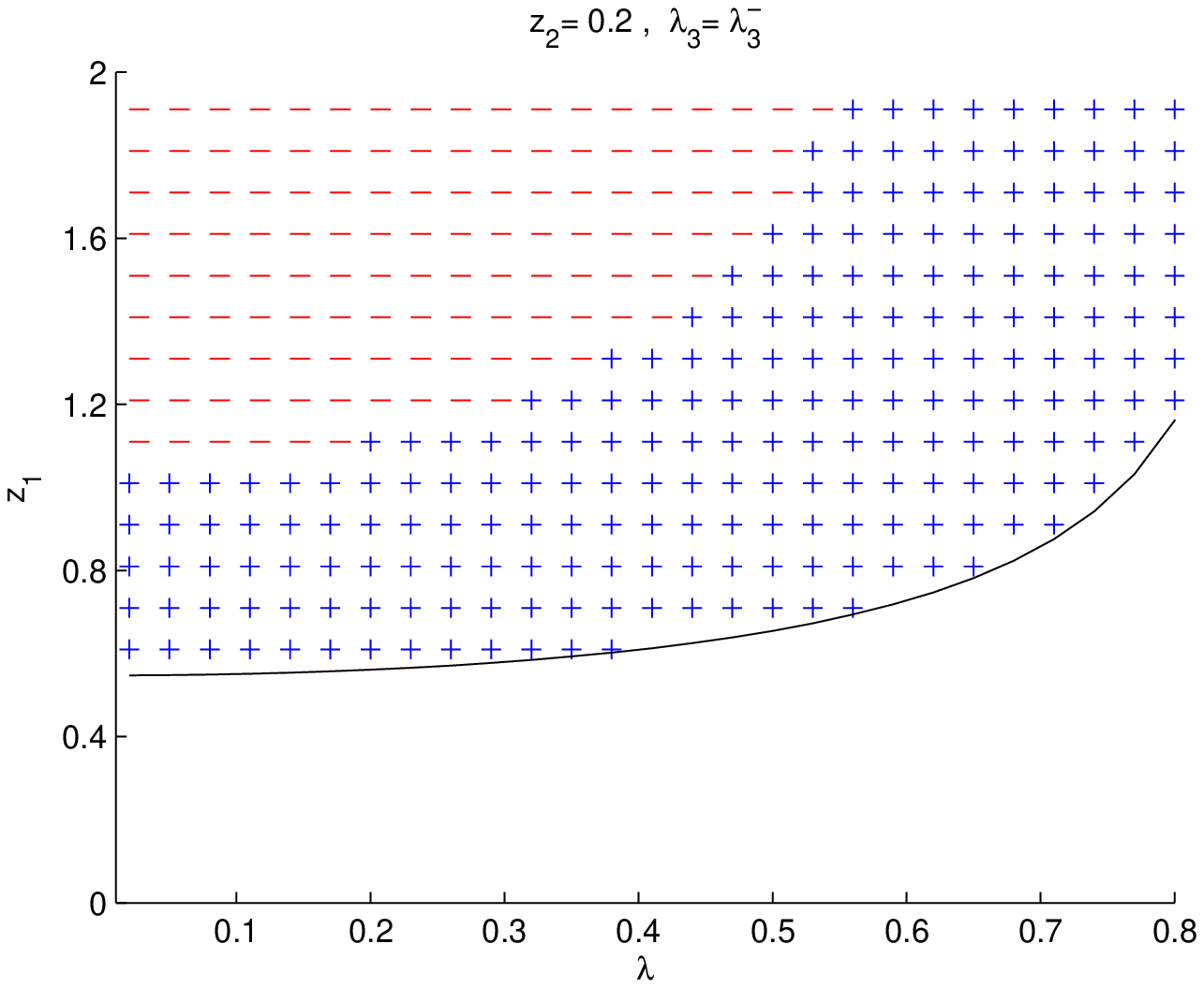} \\
                   \textit{a} & \textit{b}\\
                   \includegraphics[scale=.4]{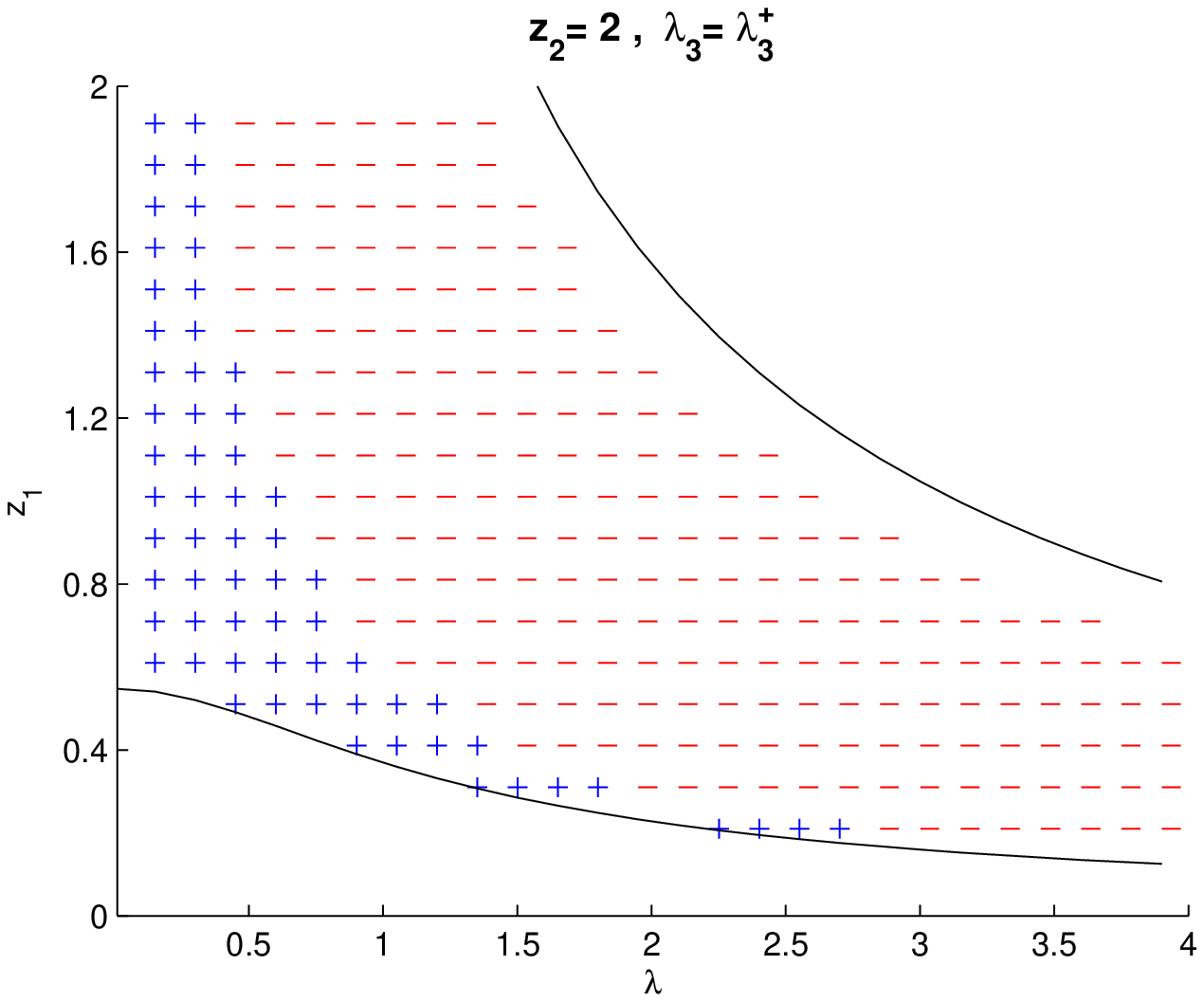} & \includegraphics[scale=.4]{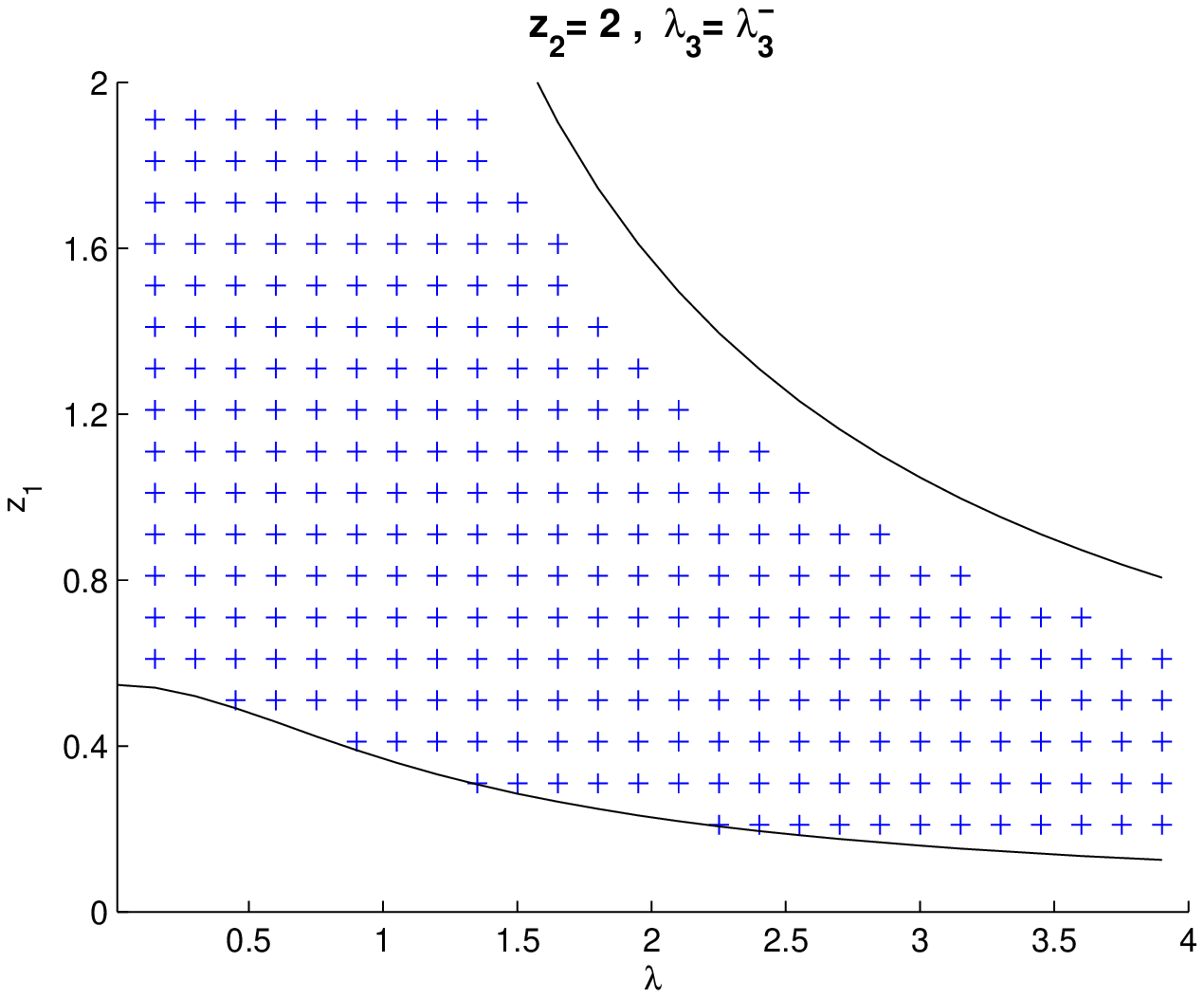} \\
                   \textit{c} & \textit{d}\\
                 \end{tabular}
 \end{center}
\caption{The plus and minus signs correspond to positive and negative tension $T$ regions, respectively.
Empty places are non-real tensions (non-real $\la_3$). The plots \textit{a}, \textit{b} are for {\it sinh}
and \textit{c}, \textit{d} are for {\it sine} cases respectively.}\label{tension}
\end{figure}

Notice that the brane tensions could be positive or negative
depending on the parameters involved ($\la$, $z_1$, $z_2$ and $L$). We may realize that we are living at $z=0$ with an
isotropic brane extension along our 4-dimensional space as in (\ref{jun2}). So the relevant brane tension
to us would be: $T$ where its sign depends on $\la$, $z_1$ and $z_2$. In figure~\ref{tension}, for one special value of $z_2$,
the positive and negative tension regions are shown for \textit{sinh} and \textit{sine} solutions, in $\la-z_1$ plane. The positive and
negative regions shrink or expand by changing the value of $z_2$.

Similar joining process should be considered for $H$ field. The Maxwell equation (\ref{Maxwell}) indicates that $h''$
is regular everywhere, on the other hand in (\ref{hprime}), $h'$ field solution admits both plus and
minus signs. Thus it should change sign while crossing $z=0$ and $z=L$. Therefore we take the plus
sign for $0 <z<L$ and minus for $-L<z<0$.  Precisely at $z=0$ and $z=L$ we take $H$ to be zero. This implies
vanishing $H$ at $z=0$ where is interpreted as the position of our 4-dimensional universe.

\newpage
\Section{Large $L$ limit}

Let us before studying the noncompact limit by sending $L$ to infinity, introduce the proper radius $\rho$ as
\bea \label{rho}
\rho=\int_0^\infty \e^{v(z)}dz
\eea
then the internal 2-dimensional metric reads as
\bea
ds_2^2= d\rho^2 + R^2(\rho)d\theta^2
\eea
where $R(\rho)=\e^{b(z(\rho))}$. Using numerical integration of (\ref{rho}), the shape of internal space is drawn for
various amounts of parameters in figure~\ref{FIG} for the \textit{sinh} case. Notice that the edges at $z=0$ and $z=L$ are the places of branes.
These are almost all possibilities that happen in the \textit{sinh} case, either in the finite $L$ or large $L$ limit. In
the rest we just concentrate on the \textit{sinh} case. For \textit{sine} case the upper limit, $L+z_1< \pi/\la$,
forbids the large $L$ limit.

Beside this numerical integration, it is worth to study the behaviors of tensions and radius of the internal space for
large $L$ limit. Firstly, for brane tensions, the results in the previous
section show that the branes at $z=0$ are untouched when $L$ is going to infinity. Thus we investigate
branes sitting at $L$ for very large $L$.

The brane tensions at large $L$  are
\bea
\kappa^2 T_{L0}|_\infty &=& \left(\frac{14}{3}\widetilde{\la}_3 +2\widetilde{\la}_4-\la \right)\e^{-v}  \nn\\
\kappa^2 T_{L4}|_\infty &=& \left(-2\widetilde{\la}_3 -2\widetilde{\la}_4-\la+2 \right)\e^{-v}  \nn\\
\kappa^2 \widetilde{T}_{L3}|_\infty &=& \left(-\frac{2}{3}\widetilde{\la}_3 +2\widetilde{\la}_4+\la \right)\e^{-v}
\eea
where $\e^{-v}$ for large $L$ is
\bea\label{eminusv}
\e^{-v} \sim A \e^{-\a L}
\eea
with $\a=(\la/4+\widetilde{\la}_3-5/4) $ and $A$ is an $L$ independent positive constant.
From equations (\ref{lambda34}), we know that $\a$ is always negative. Thus
all tensions goes to infinity at asymptotic distances.

\begin{figure}
 \begin{center}
     \begin{tabular}{c c}
       \includegraphics[scale=.4]{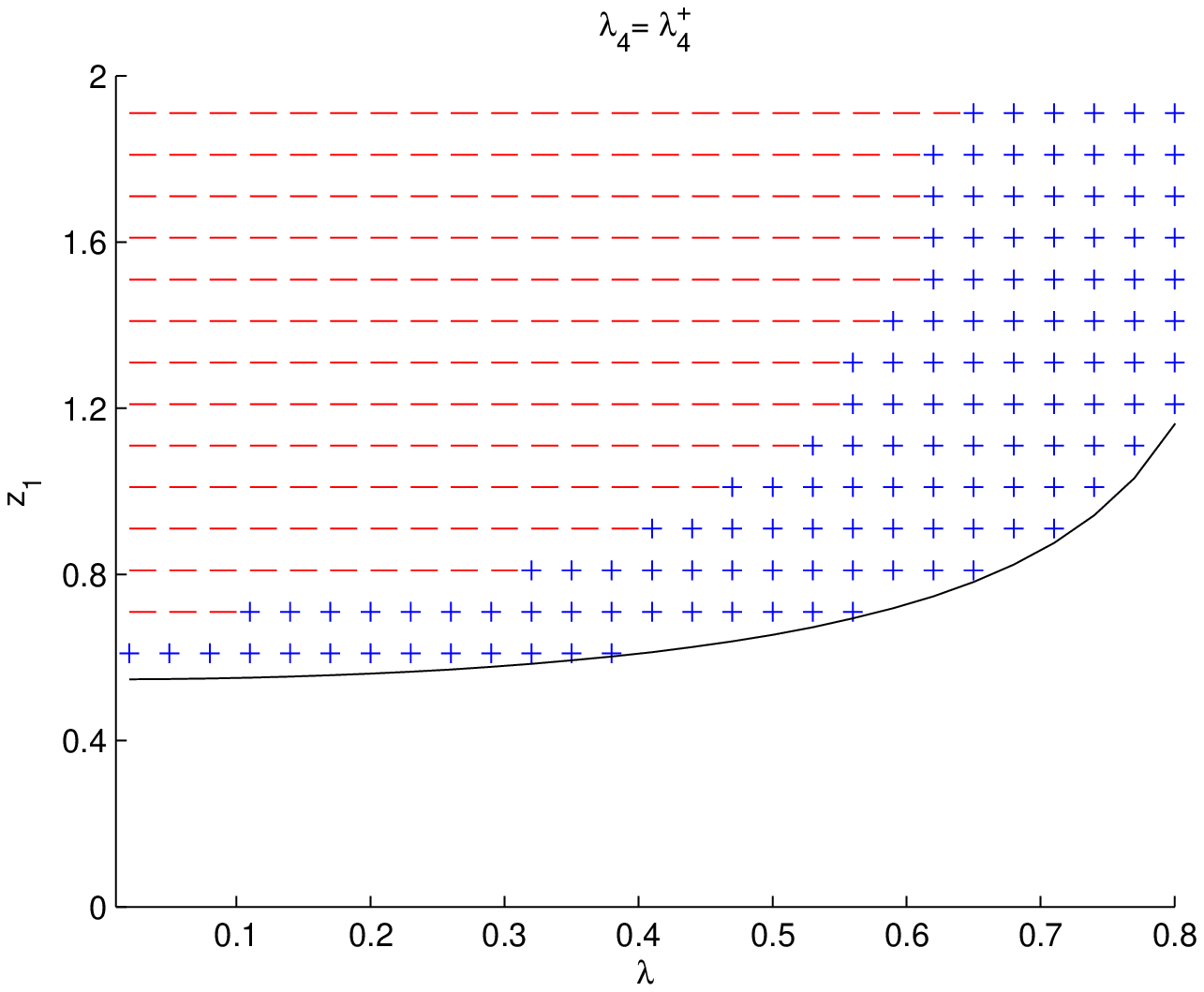} & \includegraphics[scale=.4]{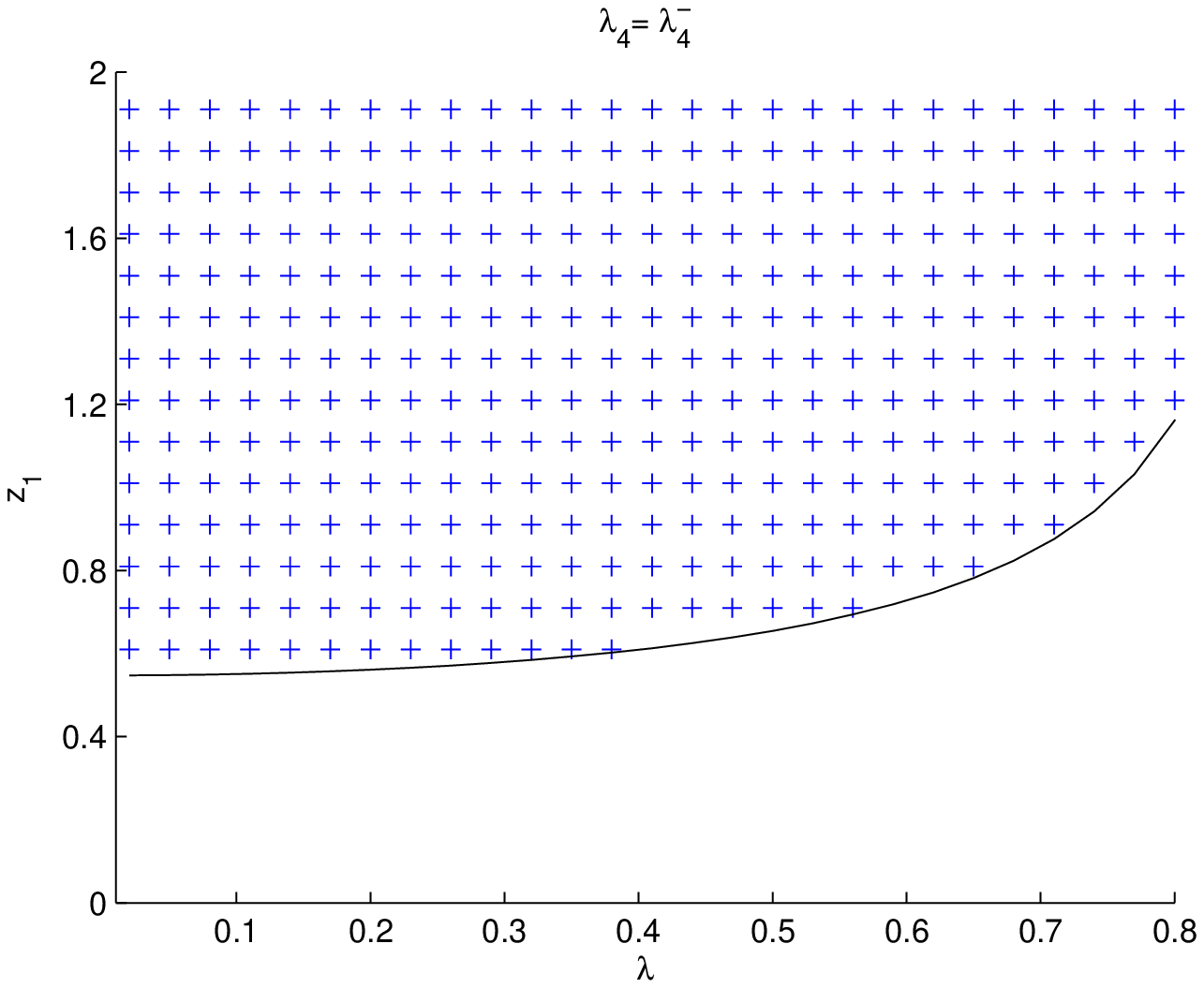} \\
       \textit{a} & \textit{b} \\
     \end{tabular}
 \end{center}
\caption{The plus and minus signs correspond to positive and
negative $\beta$ regions in the \textit{sinh} case, respectively.}\label{beta}
\end{figure}

Now look at (\ref{eminusv}), $\rho$ can be found for large $z$ as,
\bea
\rho \sim \frac{1}{A} \int_0^z \e^{\a z}dz = \frac{1}{A\a}\left(\e^{\a z}-1\right)\;.
\eea
Since $\a$ is negative, as $z$ goes to infinity $\rho$ approaches to $-1/(A\a)$ and for the radius we have,
\bea
R_L \sim (1+A\a\rho)^{-\frac{\b}{\a}}
\eea
where $\b=(\widetilde{\la}_4+\la/4+1/4)$. Thus as $z$ goes to infinity, for negative $\b$, $R_L$ diverges and
we have a noncompact space, while for nonnegative $\b$ the radius approaches to zero and a compact space is obtained
(see figures~\ref{FIG},~\ref{beta}).

\Section{Time warp consideration}

Let us look at the null energy condition which can be stated as follows\cite{Gubser:2008gr},
\bea
\tilde{T}_{MN}\xi^M\xi^N \ge 0
\eea
where $\tilde{T}_{MN}$ is constructed from energy-momentum tensor as
$\tilde{T}_{MN}=T_{MN}-\frac{1}{d-2}g_{MN}T^L_L$ in a d-dimensional space. Then by the Einstein equation it leads to,
\bea
R_{MN}\xi^M\xi^N \ge 0
\eea
for any time-like or null vector $\xi^M$.

Before checking out this condition in our case, we remind a related no-go theorem in \cite{Gubser:2008gr}, which states
that for a class of solutions named `time warp' the null energy condition can not be satisfied for compact extra
dimensions. The time warp solutions are introduced as,
\bea
ds_d^2=\e^{2A(y)}\left[-h(y)dt^2+d\vec{x}^2 \right]+\e^{2B(y)}d\tilde{s}^2_{d-4}
\eea
where $y$ denotes the compact coordinates. The above metric covers our solution with $A=a$ and $h=\exp(2w-2a)$.
With this ansatz, the null energy condition gives,
\bea \label{nullcond}
4h^2\e^{2B}\left(-R^0_0+R^1_1 \right) = -3\tilde{g}^{mn}\p_m h\p_n h + \tilde{\Box}\left(h^2\right)
+h \tilde{g}^{mn}\p_m h \p_n\left( 8 A +2(d-6) B \right)  \ge 0\nn\\
\eea
where $m$ and $n$ are extra directions indices. For $d \ne 6$ one can set $B=\frac{4}{6-d}A$ using the gauge freedom
in $y$ coordinate, then,
\bea
-3\tilde{g}^{mn}\p_m h\p_n h + \tilde{\Box}\left(h^2\right) \ge 0
\eea
Integrating over the compact extra dimensions implies $h$ to be a constant.

Notice that this argument is valid only for $d \ne 6$. For our metric in (\ref{metric}) which is in $d=6$  we find,
\bea
\e^{2v}\left(-R^0_0+R^1_1 \right) = w''-a''+(w'-a')^2+(b'-v')(w'-a')+4a'(w'-a') \ge 0
\eea
which can be converted to the form of (\ref{nullcond}) for $d=6$ with $h=exp(2w-2a)$ and $b=v$. We have already
chosen a gauge freedom in (\ref{gauge}) by which we can scape the no-go theorem. Plugging (\ref{gauge}) in the
above inequality one finds the following simple constraint,
\bea
w''-a'' \ge 0
\eea
On the other hand,
\bea
w''-a''&=& x''  \nn\\
&=& \la^2 \left(-1+\coth^2(\la( z+z_1)) \right) \ge 0
\eea
which is always true (for $sin$ and $linear$ case one can send $\la$ to $i\la$ and zero, respectively, which both
satisfy the inequality). This shows that we have constructed a solution which satisfies the energy constraint and
escapes the no-go theorem, even in the compact case. There is no contradiction here, since the no-go theorem is valid
for $d \ne 6$ and we have a counterexample for $d=6$.

\Section{Effective 4-dimensional Planck mass}

In the usual extra dimensional theories,
effective 4D theory is obtained via integrating over the extra
dimensions and interpreting the higher dimensional M-Planck
multiplied by the volume of extra dimension as the effective 4D
M-Planck. However, the warp factor of time is different from the warp
factor of space in here, so we should change the usual procedure.

Let us decompose the 6-dimensional Ricci scalar to the 4-dimensional one in the action as,
\begin{eqnarray}
  S_6 &=& M_{(6)}^4\int\sqrt{-G}\;R^{(6)}\;d^{6}x \nn\\
   &=&
M_{(6)}^4\int{d^{4}x\sqrt{-g}\bigg{(}-R_{00}^{(4)}\int{d\theta d\eta\sqrt{G}e^{-2w}}+
\delta^{ij}R_{ij}^{(4)}\int{d\theta d\eta\sqrt{G}e^{-2a}}\bigg{)}}
\end{eqnarray}
where $R^{(6)}$ is the 6D Ricci scalar, $g$ is the determinant of
the flat metric of 4D theory, G is the determinant of 6D theory and
the 6D Planck-mass is, $M_{(6)}^4=\frac{1}{2\kappa^2}$. We require
that:
\be
\int{d\theta d\eta\sqrt{G}e^{-2w}}=\int{d\theta d\eta\sqrt{G}e^{-2a}}=: V
\ee
where the integration is over the range of $\eta$. Now we define the
4D Planck-Mass as:
\be
M_{(4)}^2=\frac{1}{\kappa^2}V
\ee
\begin{eqnarray}
\int{d\theta d\eta\sqrt{G}e^{-2w}} &=& \int{d\theta d\eta \;e^{2y-x}\;e^{-2(\lambda_3+\lambda_4)\eta}} \nn\\
\label{mplanck1}   &=& L_{\theta} \frac{\kappa^3 \widetilde{q}}{\lambda_1 \widetilde{g}^2}
   \int_{-L}^{L}{dz \;\frac{\sinh(\lambda(|z|+z_1)}{\cosh^2(|z|-z_2)}e^{-2(\widetilde{\lambda}_3
   +\widetilde{\lambda}_4)|z|}} \nn\\
   &=:&L_{\theta} \frac{\kappa^3 \widetilde{q}}{\lambda_1 \widetilde{g}^2} V_1 \\
\int{d\theta d\eta\sqrt{G}e^{-2a}} &=& \int{d\theta d\eta \;e^{2y}\;e^{\frac{8}{3}\lambda_3\eta}} \nn\\
\label{mplanck2}   &=& L_{\theta} \frac{\kappa^2 }{\lambda_2 \widetilde{g}^2}\int_{-L}^{L}{dz
   \;\frac{e^{\frac{8}{3}\widetilde{\lambda}_3|z|}}{\cosh^2(|z|-z_2)}}  \nn\\
   &=:& L_{\theta} \frac{\kappa^2 }{\lambda_2 \widetilde{g}^2} V_2
\end{eqnarray}
Equating (\ref{mplanck1}) and (\ref{mplanck2}) fixes one parameter
say $\kappa\widetilde{q}$,
\bea
\kappa\widetilde{q} = \la \frac{V_2}{V_1}
\eea

\begin{figure}
 \begin{center}
  \includegraphics[scale=.5]{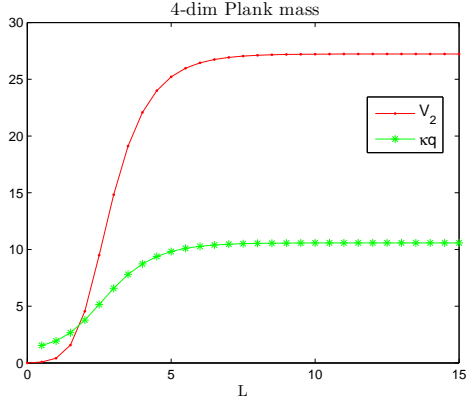}
 \end{center}
\caption{The planck mass $M_P= \frac{L_{\theta}}{\lambda_2
\widetilde{g}^2} V_2$ and $\kappa\widetilde{q}$ as functions of $L$,
for $\la=0.3$, $z_1 =0.7$ and $z_2 = 2$.}\label{Plank}
\end{figure}

Finally we find the effective
4-dimensional theory as,
\bea
S^{(4)} &=& \frac{1}{\kappa^2}V \int{d^{4}x\sqrt{-g}R^{(4)}}
\eea
with $V$ given in (\ref{mplanck2}).

Notice that our model starts with an asymmetrical spacetime due to the presence of the 3-form field $H$, however
at the end by a fine tuning of $\widetilde{q}$ which is the charge of $H$, one can reach to an effective 4-dimensional
symmetric gravity.

It is worth to consider the large $L$ limit which correspond to the case that the extra dimension is not compact and
the branes at $L$ are sending to infinity. The integrals in (\ref{mplanck1}) and (\ref{mplanck2}) remain finite for
$L \goes \infty$ which gives us a finite effective 4-dimensional Planck mass (see figure~\ref{Plank}).

\Section{Conclusion}

We have solved the static equations of motion for 6-dimensional
Salam-Sezgin model in the presence of 3-form field $H$ which provides a 4-dimensional compactification.
To find out a global solution over the compact manifold,
we consider different patches and join them with the Israel junction conditions which can be satisfied
with inserting some branes at the junctions. These conditions also fix some integration constants. More explicitly,
we have considered the compact space with angular coordinate $\theta$,
and radial coordinate $z$ where the space is defined to be periodic with fundamental region $z\in [-L,L]$ and even under
$z\goes -z$. This gives the torus topology. Then to satisfy the Israel conditions, 3 and 4 branes are
inserted at $z=0$ such that they are extended along our 4 dimensional space-time
and 4 brane wrapped and 3 branes smeared over the $\theta$ circle. The situation is the same at $z=L$ except
that we need to add some 0-branes smeared over the 4 dimensional worldvolume of the 4-brane. We may consider $z=0$ where
our brane-universe sits.

We have studied the solution behaviors in different regions of independent parameters and specially for large $L$ limit
we found that in some cases the internal radius of $\theta$ circle shrinks
and changes the topology from torus to sphere.

The asymmetry in space and time is due to the presence of the $H$ field. This kind of warping with different time and
space warp factors are recently studied in \cite{Gubser:2008gr} and called `time warp'
compactification. It is known that this compactification violates the null energy condition in $d\ne 6$
dimensions \cite{Gubser:2008gr}. However our compactification which is of course for $d=6$, shows that
the null energy condition is satisfied with a time warp compact space. In section 5, We tried to show why this happens.

Our branes configuration makes it possible to have a 4-dimensional symmetric
space at $z=0$. This can be supplemented with the fact that $H=0$ at $z=0$. Indeed $H$ is
discontinuous at this position, and changes the sign while crossing the brane. The mean value of $H$ would be zero.
There is another view in which the $H$
field exponentially vanishing at the other end, $z=L$, for very large $L$.
This enables us to reverse the situation by putting
0-branes at $z=0$ and find a symmetric space-time at $z=L$ for large $L$ where $H$ vanishes and branes preserve
the lorentz symmetry.

Another important issue is introducing an effective 4-dim Planck mass. We have done it by firstly expanding the
6-dimensional gravity action and then integrate out the extra dimensions. Since the solution has two different warp
factors for time and space, we encounter with two different integrations. Equating these two integrals we fix the
charge $q$ of the $H$ field and we can factor out integrals over the internal space and find the 4-dimensional
Planck mass.

This model is restricted to a static solution, the next development should be a dynamic solution in which all fields
would be time dependent. This is consistent with the presence of $H$ and would be important if one is interested
in finding cosmological application of this model. The stability of this model should be checked and may stabilize
some parameters (work in progress).

\newpage
\begin{center}
{\large {\bf Acknowledgements}}
\end{center}

We wish to thank M. Alishahiha, A. Davody, G.W. Gibbons, M. M.
Sheikh-Jabbari for useful conversations. Our special thanks are
devoted to H. Firouzjahi for really useful discussions and
comments in this issue. This work is partially supported by Iranian TWAS at ISMO.


\end{document}